# A passive, battery-free flexible metasurface for implantable millimetre-wave radar glucose sensing

*Kavya Saravanan [a], Rohini Ramachandran [a], Jean Pierre Ndabakuranye [a], Isabelle Shearer [b], Gursimran Singh [a], Bobbi Fleiss [b], Omid Kavehei [c,d], Akram Al-Hourani [e], Francisco J. Tovar Lopez [a,f], Margaret Lech [a], Arman Ahnood [a]**

[a] *School of Engineering, RMIT University, Melbourne, VIC 3000, Australia.*
[b] *School of Health and Biomedical Sciences, RMIT University, Melbourne, VIC 3000, Australia.*
[c] *School of Biomedical Engineering, The University of Sydney, Sydney, NSW 2006, Australia*
[d] *Sydney Nano Institute, The University of Sydney, Sydney NSW 2006, Australia*
[e] *Electrical and Electronic Engineering, University of Melbourne, Parkville, VIC 3010, Australia*
[f] *Department of Diabetes, School of Translational Medicine, Faculty of Medicine, Nursing and Health Sciences, Monash University, Melbourne, Victoria, Australia*

* *Arman.ahnood@rmit.edu.au*

Continuous glucose monitors (CGM) break the skin barrier and require periodic sensor replacement. We investigated whether a passive, battery-free flexible metasurface, implanted beneath tissue and read by a commercial 60 GHz radar, can track blood glucose levels. The metasurface contained no electronics or reagents. We fitted a conical helix to the in-phase and quadrature radar signals across the frequency sweep and used its mean radius (the spiral radius) as the glucose-dependent feature. We tested glucose concentrations of 25 to 400 mg/dL in saline solution with and without the metasurface, and 196 to 596 mg/dL of total glucose in porcine whole blood beneath ex vivo porcine tissue and in a murine cadaver model with pulsatile blood flow. The metasurface was essential for achieving a monotonic sensor response; in saline, the coefficient of determination ($R^2$) of a logarithmic fit increased from 0.18 without the metasurface to 0.90 with it. In a murine (rat) cadaver model, the signal's spiral radius increased monotonically with glucose concentration despite pulsatile flow ($R^2$= 0.91), and this monotonic behavior persisted through up to 6 mm of tissue. Although sensitivity decreased at higher glucose levels, each discrete concentration step remained clearly detectable. These results demonstrate that a shallow, passive implant paired with a commercial radar can effectively track blood glucose without requiring onboard electronics or batteries.

**Highlights**

- Passive flexible metasurface sensor is interrogated by 60 GHz radar.
- 3D spiral fitting of I/Q radar signals extracts blood glucose features.
- Metasurface integration improves saline linearity from 0.18 to 0.90.
- Murine cadaver model tracks glucose from 196 to 596 mg/dL ($R^2$ = 0.91).

- Link budget validates sub-1 mm transcutaneous sensing capabilities. (68 char)

## 1. Introduction

Diabetes management depends on frequent glucose measurement. Clinical practice relies on enzymatic tests, administered either by finger-prick or through subcutaneous continuous glucose monitoring (CGM) systems [1-3]. Although CGM has improved glycaemic control, the sensor filament remains inserted through the skin barrier, and each periodic sensor replacement requires a new filament placement [4]. Some users have also reported discomfort and skin reactions [5]. A glucose sensor that needs no reagents, batteries or replacement would remove these burdens [6]. Such a device has remained an open goal in biomedical engineering for more than three decades [7].

Millimetre-wave (mmWave) sensing at 30 ~ 300 GHz offers a reagent-free alternative. Glucose alters the complex permittivity of blood by perturbing water's hydrogen-bonding network and contributing its own dielectric response [8, 9]. Transmission and reflection measurements at mmWave frequencies respond to these dielectric changes [10, 11]. Compact, industry-standard 60 GHz radar modules make portable readers feasible [12, 13]. Indeed, miniature frequency-modulated continuous-wave (FMCW) radars at 60 and 77 GHz have already detected glucose-dependent changes in liquid samples [13-15].

Three limitations have kept these demonstrations from medical use. Firstly, glucose changes permittivity only slightly at clinically relevant concentrations, whereas tissue is heterogeneous, dispersive and ionically conductive [8, 11]. In addition, water-rich tissue attenuates 60 GHz radiation by approximately 21 dB/mm in muscle [16], albeit this varies with tissue type and state. Indeed, glucose has been measured at 60 GHz through thin animal tissue, such as porcine ear, only in proof-of-concept studies [17]. Finally, most demonstrations rely on benchtop instruments, such as vector network analysers, to measure broadband resonance shifts [10, 11, 18].

Metasurfaces may address these limitations. Metasurfaces are planar arrays of sub-wavelength unit cells that shape the amplitude and phase of electromagnetic fields [19, 20]. Passive metamaterials have also enabled contactless cardiopulmonary monitoring from everyday surfaces such as textiles [21]. For glucose sensing, a transmissive metasurface placed between a 60 GHz radar and a water-filled phantom increased the near-field power density inside the phantom by 11 dB [12]. However, that study placed the metasurface outside the body and resolved glucose in steps of 15 mg/mL (1500 mg/dL), far outside physiologically relevant changes [12]. Recently, a passive hydrogel-loaded resonator read by a commercial 60 GHz radar transduced near-field pH changes [22]. To our knowledge, no group has evaluated a passive metasurface implanted in tissue and read by a compact radar for glucose sensing in biological media.

Here, we present a passive, battery-free flexible metasurface for implantable mmWave glucose sensing, read by a compact 60 GHz FMCW radar (Fig. 1a). In the envisaged system, a wearable reader interrogates the implanted metasurface through the skin (Fig. 1b). We fabricated the metasurface on a 70 μm polyimide substrate by standard printed-circuit manufacturing; it contains no electronics or reagents (Fig. 1c). Commercial FMCW radars are designed for range estimation and positioning. They use range–Doppler maps to track targets in the far field, usually metres away. Here, the antennas, the overlying tissue, the sample and the metasurface all lie within a single range cell of 46.9 mm, so range-based processing cannot resolve them.

We therefore use the radar (Fig. 1d) as a coherent swept-frequency reader. The 3.2 GHz sweep (60.1–63.3 GHz) of this industry-standard miniature module is too narrow to track a broadband resonance shift, so we do not attempt to measure the resonance frequency. Instead, the radar records the in-phase (I) and quadrature (Q) signals at 128 frequency steps across the sweep. These samples describe the complex reflection response of everything within the range cell across the 3.2 GHz band, analogous to a network analyser. Plotted against analogue-to-digital converter (ADC) sample number, the I and Q signals trace a three-dimensional spiral (Fig. 1e). We fit a conical helix to this spiral by Markov chain Monte Carlo (MCMC) sampling [23, 24]. The radius of the helix is a signal amplitude, and therefore its square is proportional to the received power. The signal also contains constant offsets, such as DC offsets and direct leakage from the transmitter, that do not change across the sweep and shift the spiral away from the origin. The fit includes a free centre that absorbs these offsets, so the radius reflects only the contributions that rotate across the sweep, including those from the sample and the metasurface (Section 2.3). The radius can therefore serve as a feature that depends on the glucose concentration of the sample.

We evaluated this approach in three studies. In the saline study, we measured glucose at 25 to 400 mg/dL in saline with and without the metasurface (Fig. 1g). In the tissue study, glucose-spiked porcine whole blood and the metasurface were encased in *ex vivo* porcine (pig) tissue (Fig. 1h). In the murine (rat) cadaver model, the same blood circulated through a silicone tube beneath the implanted metasurface (Fig. 1f). These studies point towards minimally invasive glucose monitoring, in which a shallow, battery-free implant is read by a wearable radar without enzymatic sensor replacement.

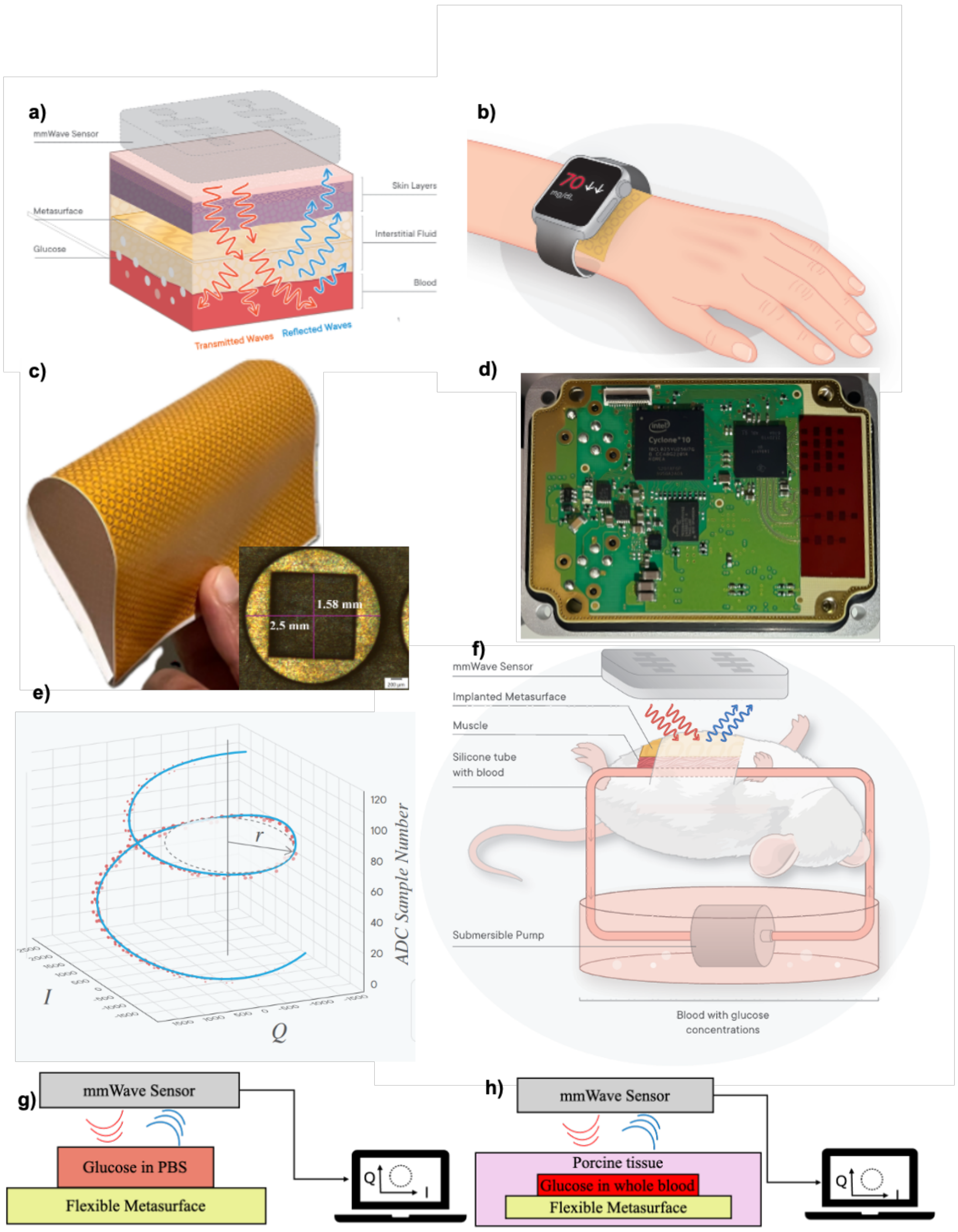

**Fig. 1. Sensing concept, components and experimental configurations.** (a) Sensing concept: a mmWave sensor transmits through the skin to a metasurface implanted above the interstitial fluid and blood, and receives the reflected waves. (b) Envisaged wearable reader, with the metasurface implanted under the skin. (c) Flexible metasurface on 70 µm polyimide. Inset: micrograph of a unit cell, a 2.5 mm metal disc with a central 1.58 mm square aperture (scale bar, 200 µm). (d) Interior of the 60 GHz FMCW radar module. (e) I and Q signals plotted against ADC sample number (red points), with the fitted spiral (blue line) and its radius r. (f) Murine cadaver model: the metasurface is implanted beneath the skin above the muscle, and glucose-spiked porcine whole blood circulates through a silicone tube driven by a submersible pump. (g) Saline study: glucose in saline solution, measured with and without the metasurface. (h) Tissue study: glucose-spiked porcine whole blood and the metasurface encased in *ex vivo* porcine tissue.

# 2. Methods

## 2.1 Radar and metasurface

We used a V-MD3 FMCW radar module (RFbeam Microwave GmbH, Switzerland; Fig. 1d), built around an IWR6843 transceiver (Texas Instruments). The module has three transmitting and four receiving patch antennas and an on-board field-programmable gate array (FPGA), and it sends data to a computer over Ethernet. We operated a single transmitter–receiver pair (TX1/RX1), with 128 samples per sweep, 64 sweeps per frame and a frame time of 130 ms.

Each sweep covered 60.1 to 63.3 GHz in 128 steps of 24.985 MHz. For each measurement, we recorded the raw I and Q samples for 30 s (231 frames) at approximately 25 °C.

We fabricated the metasurface by standard printed-circuit-board manufacturing (JLCPCB, China), with gold-plated copper unit cells on a 70 µm polyimide substrate (Fig. 1c). Each unit cell was a 2.5 mm metal disc with a central 1.58 mm square aperture, on a 2.8 mm lattice. We simulated its transmission coefficient, $S_{12}$, from 50 to 70 GHz with the frequency-domain solver of CST Studio Suite 2025, using two waveguide ports 7.5 mm from the metasurface (Section S1.1). For saline and tissue studies, the metasurface was positioned beneath the Petri dish, whereas in the murine cadaver model, it was placed subcutaneously above the thoracic muscle.

### 2.2 Samples and models

*Saline study.* The glucose (Sigma-Aldrich) was dissolved in saline (4 g/L NaCl in deionised water) and prepared at concentrations of 25, 50, 100, 200, and 400 mg/dL by serial dilution of a concentrated stock. Each 30 mL sample was mixed on a magnetic stirrer and transferred to the lid of a 90 mm polystyrene Petri dish immediately before measurement, giving a depth of about 5 mm. In each case, the concentration was measured with and without the metasurface beneath the dish, with the radar resting on the dish rim (Fig. 1g).

*Whole blood.* Porcine blood was collected from an abattoir (JBS Pork Pty Ltd, Laverton North, Australia) immediately after slaughter, serving as a practical surrogate for human blood. We anticoagulated it with crystalline trisodium citrate (3.2 g/L; Sigma-Aldrich). Crystalline citrate avoided diluting the sample. The blood was refrigerated at approximately 5 °C, and all measurements were completed within 12 h of collection. Its endogenous glucose concentration, measured with a Contour Next glucometer (Ascensia Diabetes Care), was 196 mg/dL (10.9 mmol/L). Glucose was added in increments of 0, 25, 50, 100, 200, and 400 mg/dL, yielding total concentrations ranging from 196 to 596 mg/dL (Table S1). Each whole-blood concentration was reported as a total value and verified with a commercial glucometer immediately prior to radar recording. Following gentle mixing, all samples were equilibrated to approximately 25 °C before measurement.

*Tissue study.* We used slabs of ex vivo porcine tissue from the topside (posterior femoral muscle; Fig. 1h). One slab lay beneath the metasurface and a 90 mm Petri dish holding 30 mL of whole blood, about 5 mm deep. One upper slab (6 mm) or two upper slabs (13 mm in total) lay on the dish in direct contact with the blood, and the radar rested on the uppermost slab. The slabs were kept at room temperature.

Murine *cadaver model.* We used an adult Wistar rat cadaver (RMIT Animal Ethics Committee approval, AEC 28453). A silicone tube entered through a small incision near the neck and passed through the thoracic cavity to an outlet near the abdomen, forming a closed flow path (Fig. 1f and Fig. 2a). The metasurface was implanted subcutaneously beneath a skin flap over the thorax and positioned directly above the muscle. The 60 GHz radar was then placed flush against the overlying skin surface (Fig. 2b), with the skin thickness above the metasurface estimated to be approximately 1 mm. To simulate physiological flow, a submersible pump (Core Electronics, Australia) was driven via pulse-width modulation using a 5 V relay and an Arduino Uno microcontroller, circulating the glucose-spiked whole blood from an external reservoir at approximately 150 pulses per minute. Glucose concentrations were evaluated in a randomized sequence; the fluidic circuit was thoroughly flushed and refilled between successive trials. Each concentration trial required approximately 5 minutes for system stabilization and execution, which included a 30-second continuous data acquisition window."

**2.3 Signal model and spiral fit**
During each sweep, the radar transmits a signal whose frequency increases linearly at a slope $S = 49.97$ MHz/μs. It samples the received signal 128 times at 2 MS/s, so sample $k$ corresponds to the frequency $f_k = f_0 + k\Delta f$, where $k$ runs from 0 to 127, $f_0 = 60.1$ GHz and $\Delta f = 24.985$ MHz. The sweep therefore covers 3.2 GHz in 64 μs.

Each reflector in front of the radar returns a copy of the transmitted signal, delayed by the round-trip time τ. Quadrature mixing gives the intermediate-frequency signal $s(k) = I(k) + jQ(k)$, to which each reflector contributes a tone at the beat frequency $S\tau$. Each frequency step advances the phase of this tone by $2\pi\Delta f\tau$, which is 0.06° per millimetre of distance, or 7.7° per millimetre over a full sweep. Each reflector therefore traces a circle in the I–Q plane, and a helix when plotted against ADC sample number (Fig. 1e). A more distant reflector turns faster, and a stronger reflection traces a wider helix.

Two reflectors can be resolved only if they are more than one range cell apart. Over one sweep, a reflector one range cell (46.9 mm) away completes one full turn. The antennas, the sample, and the metasurface all lie within a single range cell, so the recorded signal is the vector sum of their contributions, together with constant complex offsets and receiver noise. Therefore, each sweep was modelled as a conical helix with a free centre as:

$$s(k) \approx (I_0 + jQ_0) + [A_0 + A_1(k - k_c)]e^{j(\omega k + \varphi)} \qquad (1)$$

Where $A_0$ is the mean radius over the 128 samples, $A_1$ is the change in radius per sample, $k_c$ is the centre sample, ω is the rotation rate, and φ is the starting phase. The free centre ($I_0$, $Q_0$) absorbs components that do not change across the sweep, such as DC offsets and leakage with near-zero delay. The radius combines every rotating contribution in the range cell. Hence, a change in any one contribution alters the radius in proportion to its amplitude and relative phase. The resonant metasurface contributes, with both amplitude and phase varying across the sweep (Section 3.2). Because the radar rested on the skin, the dish rim or the tissue, the sample also lay close to or within the reactive near field of the antennas, so the measured signal may combine reflection and antenna loading (Section S1.4).

Raw I/Q data were processed using MATLAB (R2024b), and the geometric model was fitted to all 128 complex samples per sweep via the emcee ensemble Markov Chain Monte Carlo (MCMC) sampler. The sampling architecture employed 50 walkers across 5,000 iterations, utilizing a burn-in phase of 100 iterations and a thinning factor of 10. MCMC estimation was selected because I/Q responses in conductive media are susceptible to non-Gaussian amplitude and phase perturbations; furthermore, stochastic sampling bypasses the local minima that typically trap deterministic least-squares algorithms. The optimized radius parameter ($A_0$) was extracted from each sweep and subsequently averaged across all sweeps within a given recording session. This ensemble-averaged mean radius, termed the spiral radius, served as the primary glucose-dependent feature.

**2.4 Response metrics**
We quantified each response using four metrics: point-to-point sensitivity, average sensitivity, coefficient of determination ($R^2$), and Pearson's $r$. Point-to-point sensitivity is the change in radius divided by the change in concentration between consecutive concentrations. Average sensitivity is the net change in radius between the lowest and highest concentrations, divided by the concentration span. $R^2$ is the coefficient of determination of a least-squares fit of radius against the logarithm of the total glucose concentration. Finally, Pearson's $r$ is the linear correlation coefficient between radius and total glucose concentration. The response was called monotonic when the radius value increased at every successive concentration. Relative changes are given with respect to the lowest concentration: 25 mg/dL in saline and the endogenous baseline in whole blood.

# 3. Results and Discussion

### 3.1 Glucose tracking under pulsatile blood flow in the murine cadaver model

In the murine *cadaver* model, the spiral radius increased monotonically with glucose concentration, from 1359 a.u. at the baseline of 196 mg/dL to 1727 a.u. at 596 mg/dL, a rise of 27.1% (Fig. 2d). Glucose-spiked porcine whole blood circulated through a silicone tube routed through the thoracic cavity (Fig. 2a). We placed the metasurface beneath the skin flap and rested the radar on the skin above it (Fig. 2b). Throughout the evaluation, the fluidic circuit maintained a consistent pulse rate of approximately 150 pulses per minute. A logarithmic regression fit across all six glucose concentrations yielded an $R^2$ value of 0.91 ($p = 0.003$), exhibiting a sensitivity of 217 a.u. (representing a 16.0% increase relative to the baseline radius) per doubling of concentration. Here, the p-value denotes the statistical significance of the regression model coefficient, confirming a strong, non-random relationship well below the standard (alpha = 0.05) threshold.

The feature extraction pipeline tracked the spiral radius directly from the full complex signal matrix, rather than relying on conventional average received power metrics. When plotted against the analog-to-digital converter (ADC) sample number, the in-phase (I) and quadrature (Q) components of each measurement sweep trace a distinct three-dimensional spiral trajectory (Fig. 2c; Section 2.3). At the hyperglycemic peak of 596 mg/dL, the extracted spiral radius consistently exceeded the baseline radius across the majority of the radar sweep duration (Fig. 2c).

Three results supported the spiral radius as the glucose feature. Firstly, radar noise changed the radius of a measured sweep by 0.23% per sweep, 70 times less than the 16.0% change per doubling of glucose, and by 0.002% after a 30 s recording (Section S2.4). Secondly, the radius ignored small movements. When we simulated shifts of up to ±0.5 mm between recordings on a measured sweep, the radius changed by 0.12%. The root-mean-square (RMS) value of I changed by 3.3%, about as much as one glucose step (2.5–6.1%; Section S2.5, Fig. S2). Finally, in the murine cadaver model, pulsatile flow appeared mainly in the spiral phase rather than in the radius (Fig. S3). Together, these results are consistent with movement changing the spiral phase and glucose changing its radius. Indeed, radar vital-sign monitoring relies on the same separation: it recovers heartbeat and breathing from the phase of 24–120 GHz FMCW signals [25]. A single I or Q signal mixes the two, whereas the spiral fit separates them.

Pulsatile flow tested this separation directly (Fig. S3). In a 30 s recording at 296.3 mg/dL, the pump pulsed the blood at about 150 pulses per minute (2.5 Hz). For each 130 ms frame, we extracted the fitted radius, R, and the average phase of the fitted spiral, $\varphi$ (Section S2.6). The power spectrum of $\varphi$ peaked at the 2.5 Hz pump frequency, and in the 1.75–3.25 Hz band $\varphi$ followed a 2.5 Hz sinusoid (Fig. S3a, c, e). Over the same 30 s, R varied by 0.8% of its mean peak-to-peak (12 a.u.), about a third of the smallest glucose step in this model (40 a.u., or 2.5%, between 296 and 396 mg/dL). The spectrum of R also peaked near 2.5 Hz, but this peak was less prominent than that of φ (Fig. S3b, d, f). In a complex signal R exp(jφ), a fractional change in R and a change in φ in radians perturb the signal equally. On this basis, the pump modulated φ by about three times as much as R. These results are consistent with the silicone tube remaining stationary while the volume of blood inside it changed with each pump pulse. A changing blood volume shifts the effective path length of the reflection, and hence its phase, more than its amplitude. In vivo, cardiac pulse and breathing would change blood volume and tissue position in similar ways, so the fitted phase could serve as a built-in motion and pulse reference, recorded alongside the glucose-dependent radius. However, this analysis used a single 30 s recording at a single concentration, and in vivo recordings would be needed to confirm that physiological motion leaves the radius unchanged.

The response was nonlinear. Point-to-point sensitivity was 2.86 a.u. per mg/dL over 196 - 221 mg/dL, peaking at 3.45 a.u. per mg/dL over 221- 247 mg/dL and fell by 87% to 0.44 a.u. per mg/dL over 396 - 596 mg/dL. This decrease suggests partial saturation of the response at higher concentrations. In the logarithmic fit, the local sensitivity was 1.59 a.u. per mg/dL at 196 mg/dL and 0.53 a.u. per mg/dL at 596 mg/dL. The tissue study showed a similar decline above about 300 mg/dL (Section 3.3). Sensitivity in the murine cadaver model rose by 21% between the first two intervals, from 2.86 to 3.45 a.u. per mg/dL. Because each concentration was recorded only once, this rise may fall within the variation between circuit refills.

The cause of the nonlinearity remains uncertain, and three explanations are possible. Firstly, the dielectric response of whole blood may itself be nonlinear. Transmission through glucose-spiked saline at 50–75 GHz varies approximately linearly with concentration [11], but no reference model yet describes the complex permittivity of glucose in blood [8]. Cells and proteins could alter this dependence. In addition, the mapping from permittivity to spiral radius may be nonlinear. The radar observes the metasurface resonance, simulated at 62.7 GHz (Section 3.2), only within its fixed 60.1–63.3 GHz sweep. Equal shifts of the resonance could therefore produce unequal changes in radius, depending on which part of it lay within the band. Finally, the spiral radius combines the metasurface contribution with the other rotating reflections in the range cell, such as those from the skin and tube. A linear change in the metasurface contribution would therefore change the radius nonlinearly, by an amount that depends on the relative phase between them. This effect depends on the geometry of each model, so it would not necessarily produce a similar decline to that seen in the tissue study. Therefore, the logarithmic fit should be considered an empirical description of the response rather than evidence for a specific mechanism.

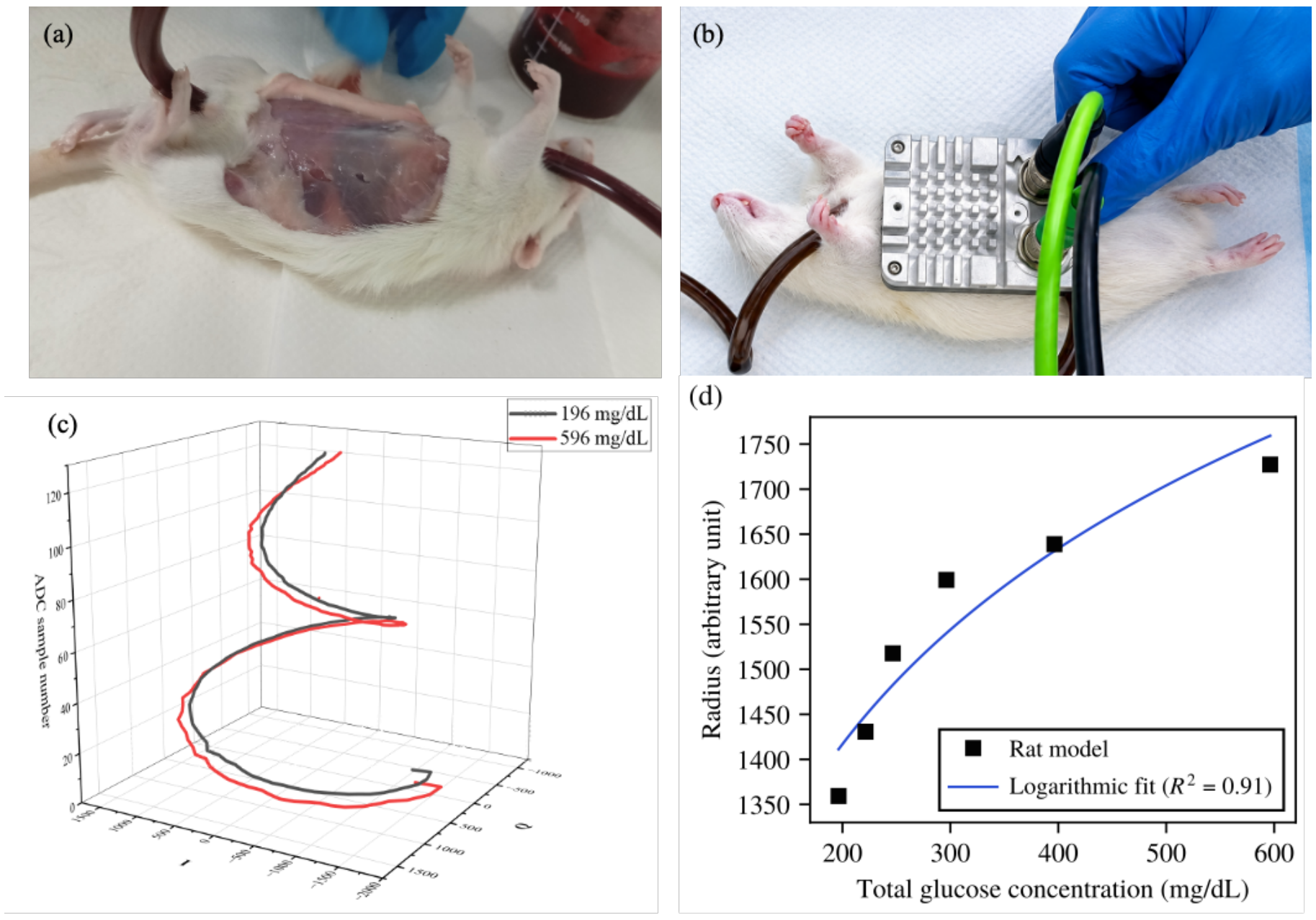

Fig. 2. **Glucose response in the *murine cadaver* model.** (a) Silicone tube carrying glucose-spiked porcine whole blood, routed through the thoracic cavity. (b) Radar positioned over the metasurface, which lies beneath the skin flap. (c) Measured I v.s. Q v.s. ADC sample trajectories at the baseline (196 mg/dL) and at 596 mg/dL. (d) Fitted spiral radius versus total glucose concentration. The line is a logarithmic fit over all six concentrations ($R^2$ = 0.91).

### 3.2 The metasurface is required for the response

A monotonic glucose response required the metasurface. Without it, the spiral radius of glucose-saline samples changed non-monotonically with concentration (Fig. 3b, c). Relative to 25 mg/dL, the radius rose by 2.6% at 50 mg/dL, fell by 12.3% and 19.8% at 100 and 200 mg/dL, and returned to - 1.5% at 400 mg/dL. With the metasurface beneath the same samples, the radius increased monotonically by 1.0%, 1.9%, 7.0% and 8.0% at 50, 100, 200 and 400 mg/dL, respectively (Fig. 3a, c). A logarithmic fit over the 25 to 400 mg/dL range gave $R^2$ = 0.90 with the metasurface and $R^2$ = 0.18 without it. The samples, radar position and signal processing were identical in both configurations, so this comparison isolates the contribution of the metasurface.

We assessed the effect of the metasurface in glucose-spiked saline without tissue, because this simpler system leaves glucose concentration as the only variable. Saline mimics the ionic properties of physiological fluids, but it is homogeneous and contains no cells or proteins. In whole blood, these constituents add their own dielectric response [26]. All whole-blood measurements then used the metasurface to test whether the response persisted under pulsatile flow in the murine cadaver model (Section 3.1) and through overlying tissue (Section 3.3).

The simulated transmission coefficient of the metasurface peaks at -39.0 dB at 62.7 GHz, within the 60.1 to 63.3 GHz band swept by the radar (Fig. 3d; Section 2.1). The frequency and strength of this resonance depend on the surrounding medium's permittivity. A small glucose-dependent change in permittivity could therefore shift the resonance, altering both the amplitude and the phase of the returned signal across the sweep. One possible explanation for the monotonic response is that the metasurface adds a distinct, permittivity-sensitive component to the received signal. The free centre of the spiral fit removes the non-rotating offsets that share its range cell (discussed further in Section 2.3), so a change in this component appears in the fitted radius.

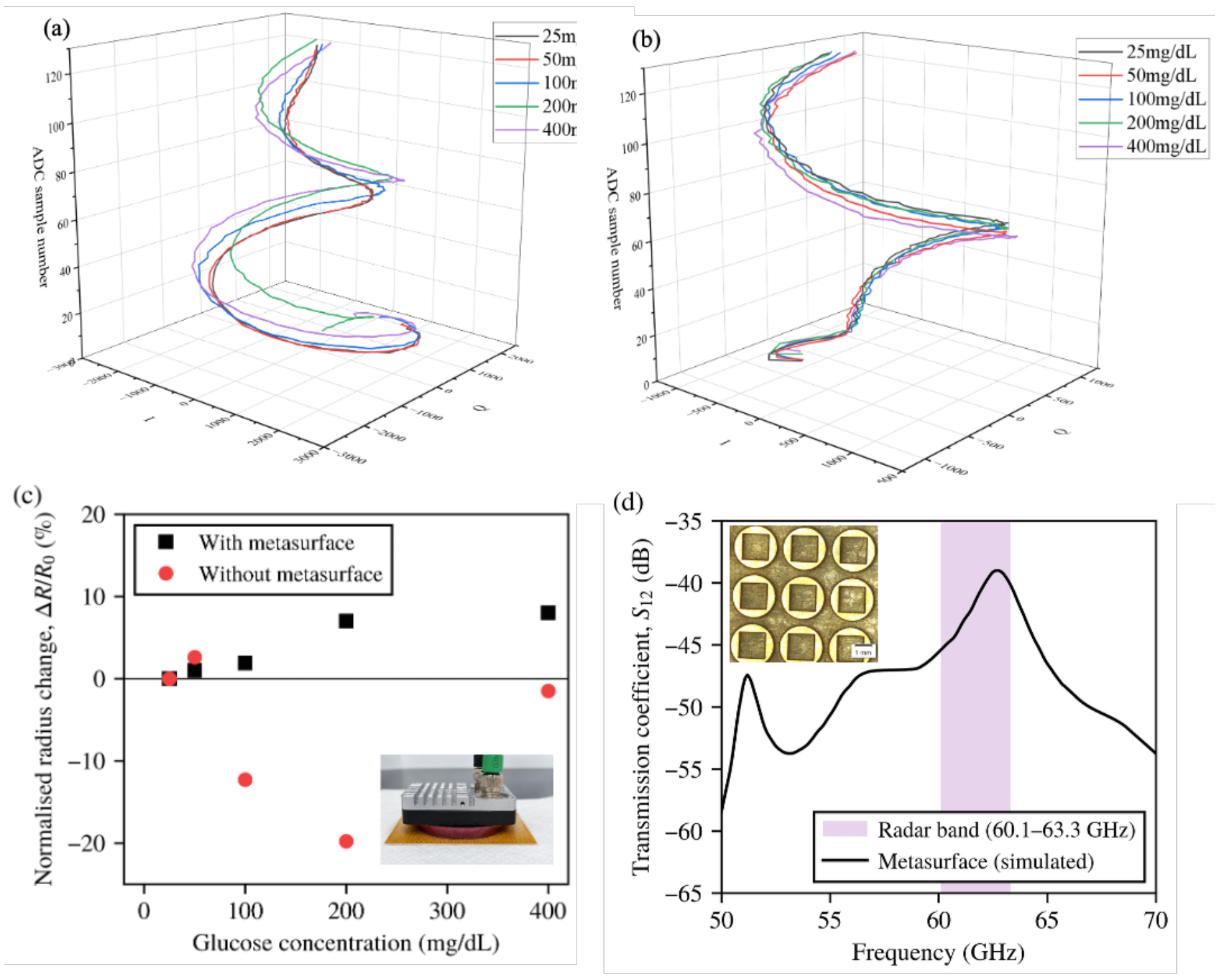

Fig. 3. **Effect of the metasurface on the glucose response.** (a, b) Three-dimensional Iv.s. Q trajectories of glucose–saline samples at 25 to 400 mg/dL, measured (a) with and (b) without the metasurface. (c) Normalised change in spiral radius, $\Delta R/R_0$, where $R_0$ is the radius at 25 mg/dL, with (black squares) and without (red circles) the metasurface. Inset: the radar resting on the rim of the Petri dish, which is placed on the metasurface. (d) Simulated transmission coefficient, $S_{12}$, of the metasurface. The shaded region indicates the radar band (60.1-63.3 GHz). Inset: photograph of the metasurface unit cells.

### 3.3 Detection through tissue and the depth limit

To test the response through overlying tissue, we encased the metasurface and a Petri dish of glucose-spiked whole blood between slabs of *ex vivo* porcine tissue, with the radar resting directly on the upper surface (Fig. 1h and Fig. 4a, inset). Through 6 mm of tissue (one upper slab), the spiral radius increased monotonically from 709.6 a.u. at the baseline of 196 mg/dL to 1119.8 a.u. at 596 mg/dL, a change of 58% (Fig. 4a). A logarithmic fit over all six concentrations gave $R^2 = 0.88$ ($p = 0.006$), with 273 a.u. (38.5%) per doubling of concentration (Fig. 4b). Point-to-point sensitivity fell from 2.27 a.u. per mg/dL over 196 to 221 mg/dL to 0.16 a.u. per mg/dL over the 396 to 596 mg/dL interval (Fig. 4c). In both the tissue and the murine cadaver model, sensitivity was higher in the three tested intervals below about 300 mg/dL. This is consistent with the logarithmic responses discussed in Section 3.1.

Adding a second upper slab (13 mm in total) reduced the consistency of the response (Fig. 4a). The radius fell from the baseline 987 a.u. to 25.3 a.u. at 221 mg/dL, then recovered to 1164 a.u. at 247 mg/dL. The same logarithmic fit gave $R^2 = 0.43$ ($p = 0.16$), compared with $R^2 = 0.88$ through 6 mm. The radius through 13 mm of tissue fell to 2.6% of its baseline value at 221 mg/dL. One possible explanation is cancellation within the range cell.
Because the measured spiral radius represents the magnitude of the vector sum of signals from all contributing reflectors, two component reflections of comparable amplitude and opposite phase can result in destructive interference. In the proximity of such a phase-cancellation boundary, minor perturbations in the phase of an individual reflection component can sharply alter the net vector magnitude. This local dip and subsequent recovery introduced an inversion in point-to-point sensitivity; specifically, the sensitivity shifted from −38.47 a.u./(mg/dL) across the 196–221 mg/dL interval to +45.19 a.u./(mg/dL) over the 221–247 mg/dL range (Fig. 4d). Between 296 and 596 mg/dL, the radius increased from 1065 to 1522 a.u. (43%), which suggests that the response to glucose persisted at higher concentrations.

Cancellation is more likely through 13 mm than through 6 mm. The additional 7 mm of tissue adds 88-297 dB of round-trip loss, depending on the fat and muscle content (Section S2.2). The blood and metasurface contributions are therefore weaker, and the slab reflections dominate the sum. When no single contribution dominates, a small change in their relative phases can cause them to nearly cancel, which is consistent with the near-zero radius at 221 mg/dL. At 6 mm, the stronger contributions from blood and the metasurface would keep the sum away from zero. However, we did not measure the individual contributions. Recordings of the slabs without blood, with replicates at 221 mg/dL, would be needed to confirm cancellation.

At an operating frequency of 60 GHz, muscle and fat tissues exhibit power attenuation rates of approximately 21 and 6 dB/mm, respectively [16] (Section S2.1). For a 6 mm heterogeneous tissue layer with equivalent fat and muscle content, the vertical round-trip propagation loss to the blood supply is approximately 168 dB (Fig. S1). While this loss falls within the maximum link budget—representing the absolute threshold between radar transmission power and receiver sensitivity—it exceeds the system's post-averaging dynamic range. However, because the precise volumetric composition of the tissue slab was not directly measured, a scenario with 6 mm of isolated fat tissue would result in only 77 dB of attenuation, which remains well within the dynamic range limits (Section S2.2). Conversely, propagating through a 13 mm tissue block would incur a round-trip signal loss ranging from 166 to 556 dB, depending on the fat-to-muscle ratio; this extreme attenuation exceeds the post-averaging dynamic range regardless of the underlying tissue composition.

In the murine cadaver model, by contrast, the metasurface lay beneath about 1 mm of skin, and the signal lost about 40 dB on its round trip (Section S2.2), well within both limits. It should be noted that these experiments show how deep the approach can reach but do not establish the optimal implantation depth.

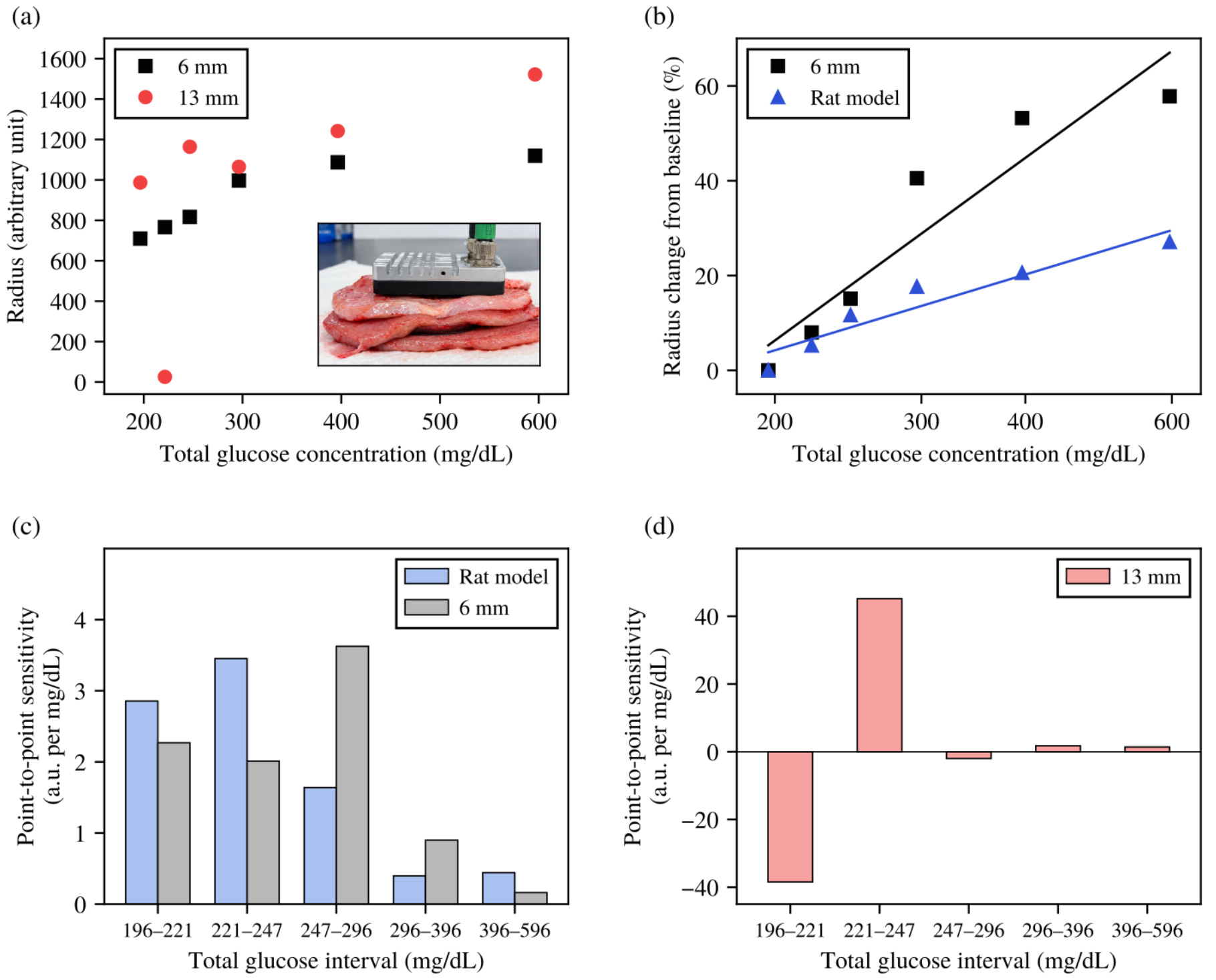

**Fig. 4. Glucose response through ex vivo porcine tissue.** (a) Spiral radius versus total glucose concentration of whole blood through 6 mm (one upper slab, black squares) and 13 mm (two upper slabs, red circles) of tissue. Inset: the 13 mm configuration, with one slab beneath the sample and metasurface, two slabs above them and the radar resting on the upper slab. (b) Radius change relative to the baseline (196 mg/dL) for 6 mm of tissue and the murine cadaver model, on a logarithmic concentration axis. Lines are logarithmic fits over all six concentrations, giving 38.5% per doubling ($R^2$ = 0.88) and 16.0% per doubling ($R^2$ = 0.91), respectively. (c) Point-to-point sensitivity of the murine cadaver model and 6 mm of tissue. (d) Point-to-point sensitivity through 13 mm of tissue.

## 3.4 Performance and limitations

The radius increased monotonically in the murine cadaver model, through 6 mm of tissue, and in saline with the metasurface, with logarithmic $R^2$ values of 0.91, 0.88, and 0.90, respectively (Table 1). In contrast, it did not increase monotonically through 13 mm of tissue or in saline without the metasurface. Average sensitivity did not follow the same order as $R^2$. Through 13 mm of tissue, it was the highest of the whole-blood configurations (1.34 a.u. per mg/dL), but $R^2$ was the lowest (0.43). In the murine cadaver model, it was the lowest of the three whole-blood expertiments (0.92 a.u. per mg/dL), but $R^2$ was the highest (0.91). Given that a large net change in radius by itself does not indicate glucose tracking, we used monotonicity and $R^2$ as the primary performance criteria.

The most similar prior work placed a transmissive metasurface between a 60 GHz radar and a water-filled phantom [12]. The received power varied by about 0.5 dB per 15 mg/mL (1500 mg/dL) step in glucose concentration, and each concentration was measured 200 times. In the present work, the spiral radius rose by 71 a.u. (5.3%) over the first 25 mg/dL step in the murine cadaver model and by 57 a.u. (8.0%) in 6 mm of ex vivo porcine tissue. This step is 60 times smaller than in [12], and the metasurface lay beneath tissue rather than outside it. The smaller step comes at a cost in repetition: we recorded each concentration once, whereas [12] repeated it 200 times, so our values carry no uncertainty estimate. The two studies also differ in sample (water versus whole blood) and readout (received power versus spiral radius). The comparison is therefore indirect, and neither study determined a detection limit.

The metasurface carries no recognition element, so it is expected to respond to changes in permittivity rather than to glucose itself. Any constituent that alters the permittivity of the blood near the metasurface could therefore change the spiral radius. Sodium chloride is one example. In a 2.46 GHz dual-resonator sensor, the output frequency shifted by 100 kHz over 0–400 mg/dL of glucose and by 145 kHz over 500–900 mg/dL of sodium chloride [27], a shift 1.45 times larger for the same 400 mg/dL span. In the present studies, only the glucose concentration varied within each series, so selectivity remains untested. A glucose-responsive hydrogel on the metasurface could provide it. A passive resonator loaded with a pH-responsive hydrogel was measured using a commercial 60 GHz FMCW radar and exhibited a monotonic response over pH 2–10 [22]. A similar layer could give the metasurface a recognition element while keeping it free of electronics. However, a glucose-responsive hydrogel has not been tested on this metasurface, and its sensitivity, response time and effect on the resonance would need to be measured.

Three key limitations qualify these results. Firstly, each concentration was recorded once, so the values in Table 1 carry no uncertainty. Furthermore, in this work we did not assess reproducibility across blood samples. Replicate sessions on independent samples would be needed to establish both. Secondly, the murine cadaver model circulated glucose-spiked porcine whole blood through a silicone tube in a pulsed circuit and therefore it did not include the exchange of glucose between blood and interstitial fluid that a subcutaneous implant would encounter *in vivo*. Further *in vivo* measurements against reference blood glucose would be needed to establish clinical accuracy. Dielectric glucose monitors have lost accuracy between controlled and daily use: in one impedance-spectroscopy study, the correlation with reference glucose fell from 0.93 under clamp conditions to 0.64 in outpatients [28]. Finally, a 0.09 dB drift in radar gain corresponds to a 1% change in spiral radius. A reference channel that senses only environmental changes could cancel such drift [27].

**Table 1. Glucose response across configurations.**

| | **Cadaver Murine model** | ***Ex vivo* porcine tissue** | | ***In vitro* saline** | |
|---|---|---|---|---|---|
| | | **6 mm** thick | **13 mm** thick | **Metasurface** | **No metasurface** |
| Metasurface to radar gap (mm) | 1 | 11 | 18 | 5 | n/a |
| Sample | Whole blood | Whole blood | Whole blood | Saline | Saline |
| Monotonic response | Yes | Yes | No | Yes | No |
| Average sensitivity (a.u. per mg/dL) | 0.92 | 1.03 | 1.34 | 0.42 | - 0.02 |
| Pearson's r (linear) | 0.90 | 0.88 | 0.64 | 0.93 | - 0.15 |
| $R^2$ (logarithmic fit) | 0.91 | 0.88 | 0.43 | 0.90 | 0.18 |

# 4. Conclusions

We developed a passive, battery-free flexible metasurface for implantable glucose sensing, read by a 60 GHz frequency-modulated continuous-wave radar. The metasurface consists of gold-plated metal unit cells on a 70 µm-thick polyimide substrate. Instead of the average received power, we tracked the radius of a spiral fitted to the in-phase and quadrature components across the 3.2 GHz sweep. This three-dimensional fit removes offsets that do not rotate over the sweep.

In the in vitro saline evaluation, the inclusion of the metasurface was critical to achieving a monotonic response. With the metasurface integrated, the spiral radius expanded by 8.0% across the 25 to 400 mg/dL concentration range ($R^2$ = 0.90)), whereas the absence of the metasurface yielded an irregular, non-monotonic response ($R^2$ = 0.18)). This specific metasurface architecture exhibits a sharp electromagnetic resonance at 62.7 GHz, which falls squarely within the radar's sweep band. Collectively, these results demonstrate that the metasurface acts as an efficient transducer, converting subtle glucose-dependent variations in complex permittivity into macroscopically measurable changes in the signal's spiral radius."

In whole blood, the spiral radius tracked glucose from 196 to 596 mg/dL. Logarithmic fits gave $R^2$ = 0.91 in the murine cadaver model with pulsatile flow and 0.88 through 6 mm of ex vivo porcine tissue. Sensitivity was highest at the low end of this range and declined at higher concentrations. Through 13 mm of ex vivo porcine tissue, the response was no longer monotonic. The depth limit of the present configuration therefore lies between 6 and 13 mm. Under about 1 mm of skin, a metasurface incurs a round-trip loss of about 40 dB. This loss lies well within both the link budget (170.5 dB) and the dynamic range after averaging (90-133 dB), supporting shallow subcutaneous placement.

Our results indicate that a passive implant, read by a commercial 60 GHz radar module, can track glucose across 196–596 mg/dL without an implanted battery or electronics. However, the murine cadaver model had no living circulation and so did not include glucose exchange between blood and interstitial fluid. In vivo measurements against reference blood glucose are now needed to establish clinical accuracy.

# 5. Acknowledgements

The authors gratefully acknowledge Sian McConchie’s contribution by providing graphic design and illustrations used in this work. This work was supported by the Australian Research Council through a Discovery Projects grant (DP230100019) and a Linkage project (LP240100251).

**A passive, battery-free flexible metasurface for implantable millimetre-wave radar glucose sensing**

**Supplementary information**

## S1 Extended methods

### S1.1 Radar and metasurface

We used a V-MD3 frequency-modulated continuous-wave (FMCW) radar module (RFbeam Microwave GmbH, Switzerland; Fig. 1d). It is built around an IWR6843 transceiver (Texas Instruments) and has three transmitting and four receiving patch antennas and an on-board field-programmable gate array. The data are transferred to a computer via an Ethernet cable for further processing. We operated it in its two-dimensional mode, which uses a single transmitter, with 128 samples per sweep, 64 sweeps per frame and a frame time of 130 ms. Each sweep covered 60.1 to 63.3 GHz in 128 steps of 24.985 MHz. We recorded the raw in-phase (I) and quadrature (Q) samples using TX1/RX1, which the module outputs as 16-bit integers from a 12-bit converter at the IWR6843, for 30 s per measurement (231 frames). All measurements were made at approximately 25 °C.

We fabricated the metasurface on a 70 µm polyimide substrate with gold-plated copper unit cells by standard printed-circuit-board manufacturing (JLCPCB, China; Fig. 1c). Each unit cell comprised a circular patch of diameter 2.5 mm with a central, metal-free square aperture of side 1.58 mm, on a 2.8 mm lattice. We simulated its transmission coefficient in CST Studio Suite 2025, which gave a peak of −39.0 dB at 62.7 GHz, inside the radar band (Fig. 3d). The $S_{12}$ response was simulated in CST Studio Suite 2025 using the Frequency Domain Solver with two waveguide ports positioned on opposite sides of the metasurface. A 7.5 mm air gap was maintained between the metasurface and the ports, with the unassigned background treated as air/free space. The simulation was performed over 50 to 70 GHz. In the saline and tissue studies, the metasurface lay beneath the Petri dish. In the murine cadaver model, it lay beneath the skin, above the thoracic muscle.

### S1.2 Samples and models

Saline: We dissolved glucose (Sigma-Aldrich) in saline (4 g/L NaCl in deionised water) and prepared concentrations of 25–400 mg/dL by serial dilution of a concentrated stock. Each 30 mL sample, about 5 mm deep, was mixed on a magnetic stirrer and transferred to the lid of a 90 mm diameter polystyrene Petri dish immediately before measurement. We measured each concentration with and without the metasurface beneath the dish, with the radar resting on the dish rim (Fig. 1g).

Whole blood: Porcine blood was collected from JBS Pork Pty Ltd (Laverton North, VIC 3026) immediately after animal sacrifice and selected as a practical surrogate for human blood owing to its biochemical similarities and ready availability. The blood was anticoagulated by directly adding crystalline trisodium citrate (S4641, $Na_3C_6H_5O_7$; Sigma-Aldrich, Australia) at 3.2 g/L. This concentration was based on the final citrate concentration obtained by mixing nine parts blood with one part 3.2% w/v citrate solution. Crystalline citrate was used instead of an aqueous formulation to avoid sample dilution and the associated artefactual reduction in glucose concentration. The resulting undiluted anticoagulated stock blood was transported immediately to the laboratory and refrigerated at approximately 5 °C. Before each measurement, the stock blood was gently vortex-mixed to minimise cellular settling, and the required aliquots were then prepared. All sample preparation and measurements were completed within 12 h of collection. Following the experiments, blood-contaminated materials were discarded in leak-proof biohazard containers, and work surfaces and blood spills were decontaminated using 1% sodium hypochlorite. Residual blood was refrigerated for no longer than seven days before disposal as biohazardous waste. The whole blood endogenous glucose concentration, measured using a Contour Next blood glucose monitor (Ascensia Diabetes Care), was 10.9 mmol/L (196 mg/dL). We added 0, 25, 50, 100, 200 and 400 mg/dL of glucose in blood, which gave total concentrations of 196 mg/dL to 596 mg/dL. We measured the glucose concentration with the glucometer immediately before each recording. Samples were mixed gently and equilibrated at approximately 25 °C before measurement.

Tissue study: We used slabs of *ex vivo* porcine tissue (muscle from posterior femoral region - main topside muscle from Pork Local Butcher chain) (Fig. 1h). One slab lay beneath the metasurface and a 90 mm Petri dish holding 30 mL of whole blood (about 5 mm deep), as in the saline study. A single upper slab (6 mm) or two upper slabs (13 mm in total) lay on the dish in direct contact with the blood, and the radar rested on the uppermost slab (Fig. 4a, inset). The slabs were kept at room temperature. For the loss calculations, we assumed equal fat and muscle content (Section S2.2).
Murine cadaver model: We used a rat cadaver (adult Wistar strain rat, RMIT ethics AEC 28453). We inserted a silicone tube through a small incision near the neck and routed it through the thoracic cavity to an outlet near the abdomen, forming a closed flow path (Fig. 1f; Fig. 2a). We implanted the metasurface beneath a skin flap over the thorax, above the muscle, and rested the radar on the skin above it (Fig. 2b). We estimated the skin above the metasurface to be about 1 mm thick. An amphibious horizontal submersible pump (Core Electronics, Australia), driven by pulse-width modulation through a 5 V relay controlled by an Arduino Uno microcontroller, circulated the glucose-spiked whole blood from a reservoir at approximately 150 pulses per minute. We measured the concentrations in randomised order. Setting up and running each concentration took about 5 min in total, of which 30 s was recorded. Between concentrations, we flushed and refilled the circuit.

Glucose totals: The glucose added to whole blood sat on top of the glucose already present, so we report every whole-blood concentration as a total (Table S1). The totals were calculated from the measured baseline and the added glucose and were confirmed with the glucometer before each recording. The baseline is the measured baseline of 196.4 mg/dL, not 0 mg/dL.

**Table S1. Added glucose and total glucose in whole blood.**

| Added (mg/dL) | Total (mmol/L) | Total (mg/dL) |
|---|---|---|
| 0 | 10.9 | 196.4 |
| 25 | 12.29 | 221.4 |
| 50 | 13.69 | 246.6 |
| 100 | 16.45 | 296.3 |
| 200 | 22.0 | 396.4 |
| 400 | 33.1 | 596.3 |

*Conversion: mg/dL = mmol/L × 18.016.*

Against total glucose, the log fits gave $R^2 = 0.91$ ($p = 0.003$) for the murine cadaver model and 0.88 ($p = 0.006$) for 6 mm tissue. The corresponding responses were 217 a.u. per doubling (16.0% of $R_0$) and 273 a.u. per doubling (38.5%), respectively. The 13 mm fit gave $R^2 = 0.43$ ($p = 0.16$). The linear fits gave $R^2 = 0.81$, 0.77 and 0.41, respectively. We report the log fit throughout. Six points per fit limit the strength of any model comparison.

## S1.3 Signal processing and analysis

We processed the raw I/Q data in MATLAB (R2024b). Plotted against analogue-to-digital converter (ADC) sample number, the I and Q samples of each sweep trace a three-dimensional spiral (Fig. 1e). We modelled this spiral as a conical helix with a free centre ($I_0$, $Q_0$), a radius that changes linearly along the sweep, a rotation rate and a starting phase. The free centre absorbs components that do not rotate over the sweep, such as DC offsets and leakage with near-zero delay, so the fitted radius excludes them. We fitted the model to each sweep by Markov chain Monte Carlo (MCMC) sampling [S1] using the *emcee* ensemble sampler, with 50 walkers, 5000 iterations, a burn-in of 100 iterations and thinning of 10. We took the mean radius over the 128 samples of each sweep and averaged it over all sweeps of the recording. This averaged mean radius is the glucose-dependent feature. We chose MCMC because I–Q responses in conductive media can contain non-Gaussian amplitude and phase variations. Sampling the parameters also avoids the local minima that can trap deterministic least-squares fitting [S1].

We quantified the response with four metrics. Point-to-point sensitivity is the change in radius divided by the change in concentration between consecutive concentrations. Average sensitivity is the net change in radius between the lowest and highest concentration, divided by the concentration span. $R^2$ is the coefficient of determination of a least-squares fit of radius against log of the total glucose concentration. Pearson's r is the linear correlation coefficient between radius and total glucose concentration. We called a response monotonic when the radius increased at every successive concentration. For whole blood, the total concentration is the added glucose plus the endogenous 196 mg/dL. For saline, the total equals the added glucose. Relative changes are given with respect to the lowest concentration (25 mg/dL in saline; the endogenous baseline in whole blood).

## S1.4 Measurement principle

The radar measures how everything in front of its antennas alters the transmitted signal. During each sweep, it raises the transmitted frequency from 60.1 to 63.3 GHz in 128 steps of $\Delta f = 24.985$ MHz. At each step, it records the returned signal as two numbers, the in-phase (I) and quadrature (Q) components. Together these give the amplitude and phase of the return. The phase measures how far the returned wave lags behind the transmitted wave. A reflection from an object at one-way distance R travels a total path of $2 \times R$ and returns after a delay of $2R/c$, where c is the speed of light. At frequency f, its phase lag is therefore 360° × f × 2R/c = 720 × f × R/c degrees. Each frequency step therefore advances the phase by 720 × Δf × R/c degrees. For Δf = 24.985 MHz, this is 0.06° per step for every millimetre of distance, or 7.7° per millimetre over a full sweep. A more distant reflector therefore turns faster in the I/Q plane. For example, a reflector 10 mm away turns by 77° over one sweep, and a reflector 46.9 mm away completes one full 360° turn.

Plotted against ADC sample number, the I and Q values of a single return therefore trace a spiral. The strength of the reflection sets the radius of the spiral, and the distance sets how fast it turns. Two returns less than 46.9 mm apart turn by less than one revolution relative to each other over a sweep, so the radar cannot separate them by range. This distance is the range cell of the radar. In this work, the antennas, sample and metasurface all lie within one range cell, so the recorded spiral is the vector sum of all their contributions. A glucose-induced change in permittivity alters this sum, and with it the fitted radius.

Close to any antenna, the electromagnetic field has two parts. The radiating part travels away from the antenna as a wave and carries energy into the surroundings. By contrast, the reactive part stays attached to the antenna and decays rapidly with distance. It stores energy and returns it to the antenna in each cycle, rather than carrying it away. This reactive part dominates within a region called the reactive near field. Its outer boundary lies at about $0.62\sqrt{(D^3/\lambda)}$ from the antenna [S2], where D is the largest dimension of the antenna and λ is the wavelength. For the four-patch antenna columns in this radar, D is 6.8–9.7 mm, or 1.4–2 wavelengths at the band centre (61.7 GHz, λ = 4.86 mm). The reactive near field therefore extends about 5.0 to 8.5 mm from the board.

Material inside the reactive near field interacts with the antenna differently from material further away. A distant object reflects part of a travelling wave back to the radar. A nearby object instead becomes part of the electrical surroundings of the antenna and alters the currents and fields on the antenna itself, an effect known as loading [S3]. In all three studies, the radar rested directly on the skin, the dish rim or the tissue, so the sample lay close to or within this region. The measured signal may therefore combine reflection and loading.

## S2 Power budget and propagation limits

Two limits decide whether the radar can detect the metasurface through a given tissue thickness. The link budget is the largest round-trip loss at which the echo stays above the receiver sensitivity. The dynamic range after averaging is the smallest amplitude change that survives receiver noise, digitisation, the fit over 128 samples and averaging over a 30 s recording. We calculated both limits and compared them with the round-trip loss of each experimental geometry (Table S3).

The link budget is 170.5 dB and the dynamic range after averaging lies between 90.3 dB (receiver noise, SNR) and 132.9 dB (quantisation noise alone, SQNR). The murine cadaver model lost 40 dB and lay inside both limits. The 6 mm tissue lost 167.7 dB, which is 2.8 dB inside the link budget and 34.8–77.4 dB beyond the dynamic range after averaging. The 13 mm tissue lost 360 dB and exceeded both limits. The dynamic range after averaging is conservative, because it is referenced to the dominant received signal rather than to the transmitted power (Section S2.3).

***S2.1 Link budget:*** The largest path loss ($L_{max}$) the transceiver can tolerate can be calculated as

$L_{max} = RP + G_{RX} - P_{RX,min}$.

Here radiated power RP = 15 dBm (typical) and 20 dBm (maximum), the receive antenna gain $G_{RX}$ = 9.5 dBi, and the receiver sensitivity $P_{RX,min}$ = −141 dBm. We assume that RP already includes the transmit antenna gain and therefore that gain is not added again. Using the maximum output power: $L_{max} = 20 + 9.5 - (-141) = 170.5$ *dB* and a typical $L_{max}$ = 165.5 dB. Both antenna gains are far-field quantities. In the reactive near field (Section S1.4), neither gain applies in full, and removing both 9.5 dB terms lowers $L_{max}$ to 146.5–151.5 dB. The measurements lie close to the boundary between the two regimes, so the far-field value is an upper bound.

We used the 60 GHz properties of skin, fat and muscle from the Gabriel model as tabulated by Wu et al. [S4]. The complex permittivity follows from the conductivity σ as

$$\varepsilon = \varepsilon' - \frac{j\sigma}{\omega\varepsilon_0}$$

$$n - j\kappa = \sqrt{\varepsilon}$$

and the power attenuation in dB per unit length is

$$\alpha = 4.343 \times \frac{4\pi\kappa}{\lambda_0}$$

with $\lambda_0$ = 4.997 mm at 60 GHz. The factor 4.343 × 4π comes from converting the power decay, $\exp\left(-\frac{4\pi\kappa z}{\lambda_0}\right)$, to decibels.

Worked example (muscle):

$$\varepsilon'' = \frac{52.80}{2\pi \times 60 \times 10^9 \times 8.854 \times 10^{-12}} = 15.82$$

$$\varepsilon = 12.86 - j15.82$$

$$n = 4.08 - j1.94$$

$$\alpha = \frac{4.343 \times 4\pi \times 1.94}{4.997} = 21.2 \frac{dB}{mm}$$

Each boundary between media with refractive indices $n_1$ and $n_2$ reflects part of the power, with a transmission loss of

$$T = -10 \log_{10}(1 - |R|^2) \text{ where } R = \frac{n_1 - n_2}{n_1 + n_2}$$

For muscle/air: $|R| = \frac{|-3.08 + j1.94|}{|5.08 - j1.94|} = \frac{3.64}{5.44} = 0.669$, so *T = 2.58 dB.*

*Table S2. Properties* of skin, fat and muscle *at 60 GHz.*

| Medium | ε | α (dB/mm) | Air-interface loss (dB) |
|---|---|---|---|
| Skin | 7.98 - 10.90j | 18.2 | 2.06 |
| Fat | 4.40 - 2.51j | 6.3 | 0.78 |
| Muscle | 12.86 -15.82j | 21.2 | 2.58 |

**S2.2 Round-trip loss**

A signal that reaches a depth and returns crosses each layer and each boundary twice:

$$L_{rt} = \Sigma_i 2T_i + \Sigma_i 2\alpha_i d_i,$$

where $d_i$ is the thickness of layer i.

For the murine cadaver model, with an estimated 1 mm of skin: $L_{rt} = 2 \times 2.06 + 2 \times 18.2 \times 1 = 40.5$ dB. With 2 mm of muscle between the skin and the blood, the round trip to the blood is 125 dB.

For 6 mm of 50% fat and 50% muscle assuming a double layer, this is $L_{rt} = 2 \times 0.78 + 2 \times 0.56 + 2 \times 6.3 \times 3 + 2 \times 21.2 \times 3 = 167.7$ dB. It should be noted that the slab composition was not measured, and it dominates this value. Composition has a large effect: 6 mm of fat alone gives 77 dB, and 6 mm of muscle alone gives 259 dB (Fig. S1, grey bars). For 13 mm, the loss ranges from 166 dB for fat alone to 556 dB for muscle alone. It should also be noted that plane-wave values are lower bounds: in the reactive near field, components with larger transverse wavenumbers decay faster inside a lossy medium.

**S2.3 Dynamic range after averaging**

The IWR6843 digitises I and Q with a 12-bit converter (±2,048 counts), and the module outputs them as 16-bit integers. Rounding each sample to a whole count adds quantisation noise of power $\Delta^2/12$ to each channel, where $\Delta = 1$ count. For a signal of amplitude A counts, the signal-to-quantisation-noise ratio of the complex sample is

$$SQNR = \frac{A^2}{\left(2 \times \frac{\Delta^2}{12}\right)} = \frac{6A^2}{\Delta^2}$$

or, in decibels, $SQNR$ = 10 $\log_{10}$ ($\frac{6A^2}{\Delta^2}$) = 10 $\log_{10}$ 6 + 20 $\log_{10}$ A. In this work, the largest recorded amplitude was A ≈ 1,311 counts. This gives

SQNR = 7.8 + 62.4 = 70.1 dB per sample.

Averaging N sweeps reduces uncorrelated noise power by a factor of N. In decibels, this is a gain of $G_{avg} = 10 \log_{10}N$. Each 30 s recording contains 231 frames of 64 sweeps, so N = 14,784 and $G_{avg}$ = 41.7 dB. Taking the mean radius over the $N_s$ = 128 samples of each sweep adds a further gain, $G_{fit} = 10 \log_{10}N_s$ = 21.1 dB. The SQNR-based dynamic range after averaging is then:

$R_{avg} = SQNR + G_{avg} + G_{fit}$ = 70.1 + 41.7 + 21.1 = 132.9 dB.

A contribution returning more than $R_{avg}$ below the dominant signal changes the averaged trace by less than the residual quantisation noise, so it cannot be resolved. This is an upper bound, because it assumes that quantisation is the only noise source and the noise is uncorrelated between sweeps.

We estimated the receiver noise at σ = 39 counts per sample in both the I and Q channels (Section S2.5), 42.6 dB above the quantisation noise. The signal-to-noise ratio per sample is therefore

$$SNR_{dB} = 10\log_{10}\left(\frac{A^2}{2\sigma^2}\right) = 10\log_{10}\left(\frac{1311^2}{2 \times 39^2}\right) = 10\log_{10}(565.00) = 27.52\ dB$$

and, with receiver noise in place of quantisation noise, the dynamic range after averaging is

$$\begin{aligned} R_{avg-SNR} &= SNR_{dB} + 10\log_{10} N_s + 10\log_{10} N \\ &= 27.52 + 10\log_{10}(128) + 10\log_{10}(14784) = 27.52 + 21.07 + 41.70 \\ &= 90.3\ dB \end{aligned}$$

The two estimates bound the dynamic range after averaging: 132.9 dB if quantisation were the only noise source, and 90.3 dB with the measured receiver noise. Fig. S1 shows this range as a shaded band, and Table S3 compares each path with both values.

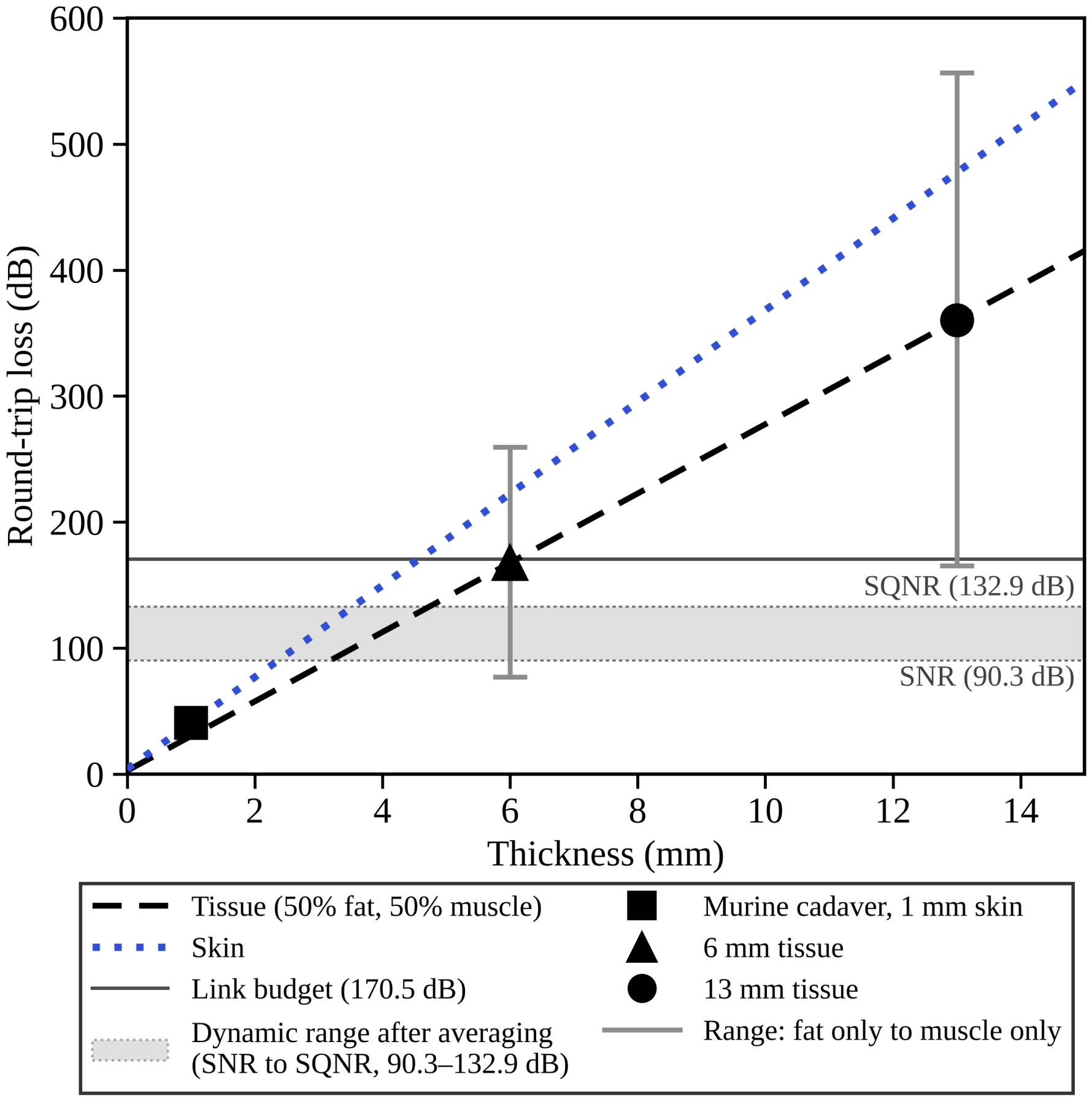

**Fig. S1. Round-trip loss of a vertical plane wave against tissue thickness at 60 GHz. The dashed and dotted curves show tissue with equal thicknesses of fat and muscle, and skin, respectively. The solid line marks the link budget (170.5 dB), and the shaded band the dynamic range after averaging for a 30 s recording, from 90.3 dB (receiver noise, SNR) to 132.9 dB (quantisation noise only, SQNR). Markers show the murine cadaver model (metasurface beneath 1 mm of skin, 40 dB) and the vertical path to the blood through 6 mm (168 dB) and 13 mm (360 dB) of tissue. Grey bars show the loss for fat alone to muscle alone.**

| **Configuration** | **Path** | **$L_{rt}$ (dB)** | **Within link budget (170.5 dB)** | **Resolvable (90.3–132.9 dB)** |
|---|---|---|---|---|
| Murine cadaver model | 1 mm skin, to metasurface | 40 | Yes | Yes |
| Murine cadaver model | 1 mm skin, 2 mm muscle, to blood | 125 | Yes | Maybe |
| Tissue, 6 mm | 50% fat / 50% muscle, to blood | 168 | Yes | Maybe |
| Tissue, 13 mm | 50% fat / 50% muscle, to blood | 360 | Maybe | No |

**Table S3. Round-trip loss of the vertical path for each experimental geometry, compared with the link budget (170.5 dB) and the dynamic range after averaging (90.3–132.9 dB). Tissue values assume equal thicknesses of fat and muscle. Tissue rows give the loss to the blood; the metasurface lay beneath a further 5 mm of blood.**

In the murine cadaver model, the metasurface lay beneath an estimated 1 mm of skin, a round trip of 40.5 dB, which is well within both limits. This supports subcutaneous placement at a depth of a few millimetres or less. The circulating blood lay beneath the thoracic wall.

For the tissue slabs, 50% fat and 50% muscle gives 168 dB through 6 mm. This lies 2.8 dB within the link budget but 34.8–77.4 dB beyond the dynamic range after averaging. Although this suggests that a signal from beneath the slab could not be resolved, the slab composition alone moves the 6 mm loss from 77 dB (fat) to 259 dB (muscle), a range that spans both limits; temperature and hydration add further uncertainty. Through 13 mm, the loss exceeds the dynamic range after averaging for any composition (166–556 dB), and the link budget unless the slab is almost entirely fat.

These estimates assume homogeneous layers, normal incidence and properties from a single dielectric model. Layered tissue can produce standing waves that change one-way transmission by up to a factor of two at particular thicknesses [S6]. That corresponds to about 6 dB over the round trip, comparable to the 2.8 dB margin of the 6 mm tissue path within the link budget and the 7.6 dB margin of the blood path in the murine cadaver model within the SQNR-based dynamic range (Table S3).

### S2.4 Why the spiral fit is not limited by noise

Each sweep contains 128 complex samples, one at each frequency step. When we fit the spiral model, all 128 samples contribute to each fitted parameter. The effect of random noise therefore shrinks as more samples are used. The smallest possible uncertainty of a fitted parameter is given by the Cramér–Rao bound. For the spiral radius A, it is

$$\frac{\sigma_A}{A} \geq \frac{1}{\sqrt{2N_s \cdot SNR}}$$

where $N_s = 128$ is the number of samples per sweep and SNR is the signal-to-noise ratio of one sample. If quantisation were the only noise source (SNR = 70.1 dB), the radius could be determined to about $2 \times 10^{-5}$ of its value, or 0.002%, from a single sweep. For the murine cadaver model, whose baseline radius was 1,359 a.u., this is about 0.03 a.u. The measured glucose response was 217 a.u. per doubling of glucose concentration, or 16.0% of the baseline radius. This is more than 7,000 times larger than the noise-limited uncertainty. Quantisation noise therefore does not limit the precision of the fitted radius. Receiver noise is larger. With the 39 counts per sample estimated in Section S2.5, the fitted radius of a measured sweep has a single-sweep uncertainty of 0.23%, or 0.002% after averaging a 30 s recording. The single-sweep value is still 70 times smaller than the response per doubling of glucose.

### S2.5 Fitted radius compared with the RMS of I or Q

We compared the fitted radius with the root-mean-square (RMS) of I and of Q as measures of a glucose-induced change. We used one measured sweep from TX1–RX1, with the metasurface 7.6 mm from the radar, inside the reactive near field (Section S1.4; Fig. S2a, b). A conical helix with a free centre fitted this sweep with $R^2 = 0.90$. We enlarged the fitted spiral by 1% and kept the measured offset and fit residual unchanged. We chose 1% because single glucose steps in the murine cadaver model changed the fitted radius by 2.5–6.1%. At 296 mg/dL, the ±15% accuracy criterion of ISO 15197 [S5] corresponds to a 1.1% change in fitted radius, and at 396 mg/dL to a 1.6% change.

A 1% increase in the fitted spiral changed the RMS of I, the RMS of Q and the fitted radius by 0.89%, 0.91% and 1.00%, respectively (Fig. S2c). The fitted radius therefore responded 10–13% more than the RMS of I or Q. We estimated the receiver noise at 39 counts per sample from sample-to-sample differences of the fit residual; the spectrum away from the signal gave the same value. At this noise level, the smallest detectable change in a single sweep was 1.36%, 1.44% and 0.97% for the RMS of I, the RMS of Q and the fitted radius, respectively (Fig. S2d). We defined this limit as three standard deviations of the difference between two recordings, estimated from 2,000 Monte Carlo trials. The 1.4–1.5 times lower limit of the fitted radius is consistent with its use of both the I and Q channels. Averaging the 14,784 sweeps of a 30 s recording lowered the three limits to 0.011%, 0.012% and 0.008%, respectively. Receiver noise therefore limits none of the three measures after averaging. However, this noise estimate comes from a single sweep; consecutive sweeps from a static target would measure the sweep-to-sweep noise and drift directly.

Displacement between recordings, in contrast, separated the RMS of I or Q from the fitted radius. At the 61.7 GHz centre frequency, a displacement of 0.1 mm in air rotates the reflection by 15°. With no glucose change, displacements distributed uniformly up to ±0.1 mm produced errors of 0.24%, 0.17% and 0.02% in the RMS of I, the RMS of Q and the fitted radius, respectively (Fig. S2e). Up to ±0.5 mm, these errors grew to 3.3%, 3.1% and 0.12%, respectively. The error in the RMS of I equalled a 1% glucose change at ±0.24 mm, whereas the error in the fitted radius was at most 0.14% up to ±0.6 mm. Over a full rotation, the RMS of I varied by 12.7% peak to peak. About half of this variation arose from the non-integer number of turns in the sweep (2.15), and most of the remainder from the fixed fit residual; a spiral with a whole number of turns and constant radius gave none. Rotation changed the fitted radius only through the fixed fit residual, because the free centre and phase of the fit follow the rotated spiral. This emulation assumes that only the spiral moves. If the fit residual moves with it, the error in the RMS of I at ±0.1 mm rises to 0.90%, and the error in the fitted radius falls to zero. These results indicate that the main advantage of the fitted radius over the RMS of I or Q is its insensitivity to displacement between recordings. Its error was 7–10 times lower at ±0.1 mm and 26–28 times lower at ±0.5 mm. Its larger response (10–13%) and lower noise limit (1.4–1.5 times) are secondary.

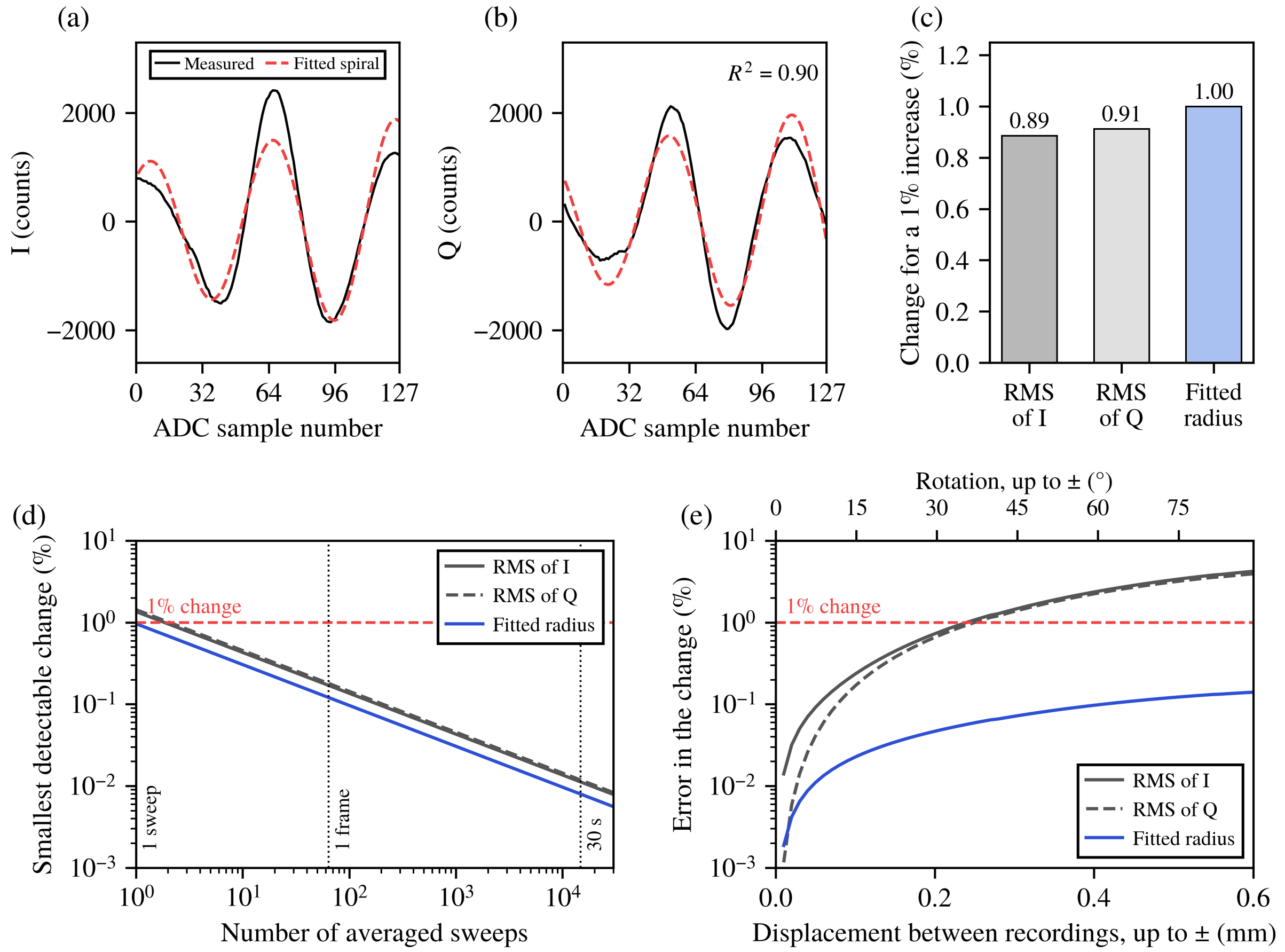

**Fig. S2. Fitted radius and RMS of I or Q for an emulated 1% change (TX1–RX1, metasurface 7.6 mm from the radar). (a) I and (b) Q of one measured sweep (black) and the fitted spiral (red dashed). (c) Change in each measure for a 1% increase in the fitted spiral. (d) Smallest detectable change, defined as three standard deviations of the difference between two recordings, against the number of averaged sweeps; receiver noise 39 counts per sample, 2,000 Monte Carlo trials. (e) Root-mean-square error in the change caused by a displacement between recordings, distributed uniformly up to the value on the lower axis, with no glucose change and only the fitted spiral moving; the upper axis gives the corresponding rotation. Red dashed lines mark a 1% change.**

## S2.6 Stability of the fitted radius under pulsatile flow

We tested whether pulsatile flow disturbs the fitted radius in a 30 s recording from the murine cadaver model at 296.3 mg/dL, with the pump running at about 150 pulses per minute (2.5 Hz). For each 130 ms frame, we extracted the fitted radius, R, and the average phase of the fitted spiral, φ (the angle of each measured I/Q sample about the fitted spiral centre, unwrapped along the sweep and averaged over the 128 ADC samples, then averaged over the sweeps of the frame; Fig. S3a, b). Over 30 s, φ drifted by about 70 mrad (4°). If displacement caused this drift, it corresponds to a change in path length of about 27 µm in air. The power spectrum of φ peaked at 2.5 Hz, the pump frequency (Fig. S3e). In the 1.75–3.25 Hz band, φ followed a 2.5 Hz sinusoid (Fig. S3c).

Over the same 30 s, R stayed between 1,566 and 1,578 a.u., a peak-to-peak variation of 0.8% of its mean of about 1,573 a.u. (Fig. S3b). This variation is about one third of the smallest glucose step in the murine cadaver model, 40 a.u. (2.5%) between 296 and 396 mg/dL. The spectrum of R also peaked near 2.5 Hz, but this peak was less prominent above the background than that of φ (Fig. S3d, f). For a complex signal R exp(jφ), a fractional change in R and a change in φ in radians perturb the signal equally, so the two can be compared directly. On this basis, the pump modulated φ about three times more than R.

These results indicate that pump pulsation appeared mainly in the phase of the reflection. This is consistent with small movements of the tube and surrounding tissue. Such movements change the path length, and hence the phase, more than the reflected amplitude [S6]. The spiral fit assigns these changes to φ and leaves R nearly unchanged. In contrast, each channel records R cos φ or R sin φ, so the RMS of I or Q combines amplitude and phase.

It should be noted that Section S2.5 showed that a rotation of the reflection changes the RMS of a single channel but not the fitted radius. Within this recording, the phase drift corresponded to about 27 µm, about nine times smaller than the ±0.24 mm at which the emulated error in the RMS of I equalled a 1% glucose change. Larger shifts between recordings, for example from repositioning the radar, would therefore affect the RMS of I or Q more than the fitted radius.

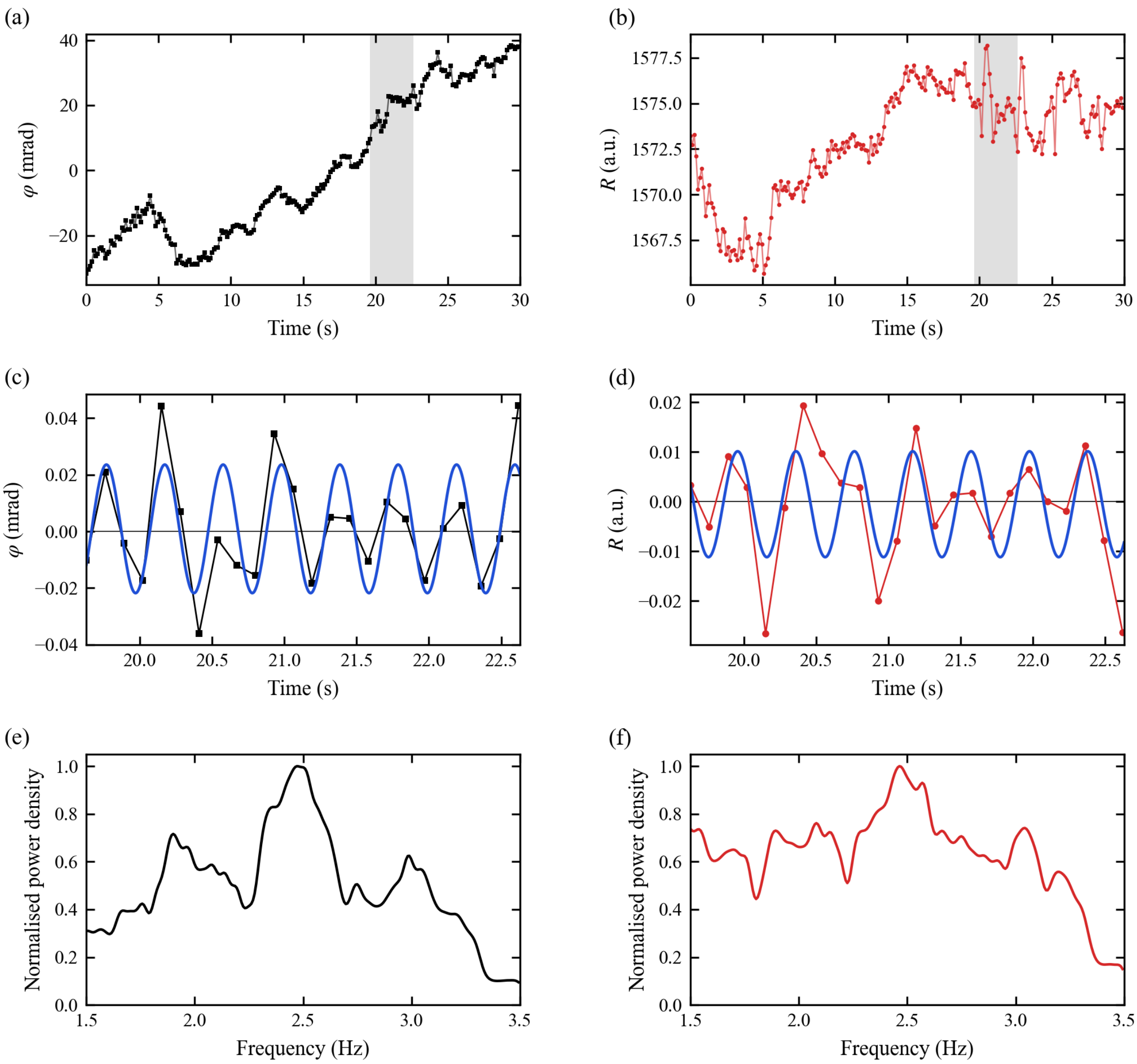

**Fig. S3. Fitted radius and phase during pulsatile flow in the murine cadaver model (296.3 mg/dL; 30 s; pump at about 150 pulses per minute, 2.5 Hz). (a) Average phase of the fitted spiral, φ, and (b) fitted radius, R, for each 130 ms frame. (c, d) The same signals band-pass filtered at 1.75–3.25 Hz over a typical 3 s window. Blue lines are fitted 2.5 Hz sinusoids. (e, f) Power spectra of φ and R, each normalised to its maximum.**